\documentclass[preprint,12pt]{elsarticle}

\usepackage{amssymb}

\newcommand{\argmin}{\mathop{\rm arg~min}\limits}

\journal{arXiv}

\begin{document}

\begin{frontmatter}



\title{Complex Valued Risk Diversification}

\author[label1]{Yusuke Uchiyama}
\author[label1]{Takanori Kadoya}
\author[label2]{Kei Nakakagawa}
\address[label1]{MAZIN, Inc., 1-60-20 Minami Otsuka, Toshima-ku, Tokyo, Japan}
\address[label2]{Nomura Asset Management Co., Ltd. 1-21-1, Nihonbashi, Chuo-ku, Tokyo, Japan}

\begin{abstract}
Risk diversification is one of the dominant concerns for portfolio managers.  Various portfolio constructions have been proposed to minimize the risk of the portfolio under some constrains including expected returns.  We propose a portfolio construction method that incorporates the complex valued principal component analysis into the risk diversification portfolio construction.  The proposed method is verified to outperform the conventional risk parity and risk diversification portfolio constructions.    
\end{abstract}


\end{frontmatter}


\section{Introduction}
\label{sec:intro}
Both individual and institutional investors are concerned with risk diversification for portfolio construction.  Portfolio managers have employed appropriate mathematical techniques to minize the risk of the portfolios, formulated as constrained nonlinear optimization problems.  Indeed, as the pioneer of quantitative finance, Markowitz proposed the mean-variance (MV) portfolio construction \cite{Markowitz}.  In the framework of the mathematical portfolio construction, the return and risk of the portfolios are defined by the mean and variance respectively, then the MV portfolio construction is determined by the minimization of the risk of the portfolio constrained with the expected return.  However, it was pointed out that the risk allocations of the MV portfolio construction are often biased \cite{Michaud}. In other words, the weight levels of particular assets are much higher than others in the MV portfolio.  \par 
In general, risk biased portfolios seem to be vulnerable to asset price change. The MV portofolio construction is thus undesireble from the point of view of risk diversification.  The risk parity (RP) portfolio construction was designed to allocate market risk equally across asset classes, including stocks, bonds, commodities, and so on \cite{Qian}.  Subsequntly, a return weighted sum of assets is introduced to the RP portfolio construction for improving its performance \cite{Baltas-Jessop-Jones-Zhang}.  Some variations of the RP portfolio construction have been proposed and verified to outperform the MV portfolio construction \cite{Lohre-Opfer-Orzag,Clarke-Silva-Thorley,Chaves-Hsu-Li-Shakernia}.  Nevertheless, the RP portfolio construction cannot fully disperse the origin of risk because almost all parts of the world mutually interact in modern society, causing entanglement of different asset classes.  \par
In the field of data science and multivariate analysis, the principal component analysis (PCA) has been developed to decompose mutually correlated data subspaces \cite{James_etal}.  The maximum risk diversification (MRD) portfolio construction utilizes the PCA to decompose and allocate the risk contribution of assets \cite{Meucci}.  Then the constrained optimization of the MRD portfolio construction is expected to design a risk allocated portfolio.  The MRD portfolio construction is also confirmed to outperform the MV portfolio construction and to be able to allocate the risk contribution of assets \cite{Meucci,Theron-Vuuren}  \par
On the other hand, in the filed of the atmospheric physics, the PCA has been utilized and extended to capture principal modes of spatiotemporal dynamics, which are known as empirical orthogonal functions (EOFs) \cite{Hannachi-Jolliffe-Stephenson}.  In practice, it is extremely difficult to investigate all the degrees of freedom of global atmospheric changes.  Thus, the method of EOFs has been employed to extract essential dynamics \cite{Smith-Reynolds-Livezey-Stokes,Newman-Compo-Alexander,Clark-Pisias-Stocker-Weaver}. \par
The conventional portfolio constructions have not considered the temporal dynamics of the portfolios despite the importance of the temporal fluctuation of assets.  In this research we incorporate dynamic effects into the MRD portfolio construction by using the method of EOFs.  The Hilbert transform is utilized to generate analytical signals from the prices of the assets.  In addition the corresponding optimization problem is presented to equalize dynamic risk allocations and then is verified to outperform the conventional methods. 
\section{Related works}
\label{sec:RW}
\subsection{Mean-variance  portfolio}
Markowitz first introduced the MV portfolio as a sophisticated method in modern portfolio theory.  In this theory, the risk of the asset is defined as the standard deviation of the return.  With this setup, a portfolio is presented by the weighted sum of the assets considered.  \par 
Given the sequence of $m$-th asset prices $\{p_t^{(m)}\}_{0{\leq}t{\leq}T}\;(1{\leq}m{\leq}M)$, the return of the asset is defined by
\begin{equation}
r_t^{(m)}=\frac{p_{t+1}^{(m)}-p_t^{(m)}}{p_t^{(m)}}.
\label{eq:RTN}
\end{equation}
Subsequently, the return of the portfolio is obtained as
\begin{equation}
R_t=\sum_{m=1}^Mw_mr_t^{(m)},
\label{eq:RTNPRT}
\end{equation}
where $\{w_m\}_{1{\leq}m{\leq}M}$ is the set of weight coefficients.  The risk of the portfolio is defined by the standard deviation of the return in Eq.~(\ref{eq:RTNPRT}).  In general risk averse investors tend to minimize the risk of their portfolios under expected returns.  This strategy is mathematically formalized by constrained quadratic programming with respect to the covariance matrix of the return of the portfolio.  \par
The expected return of the portfolio in Eq.~(\ref{eq:RTNPRT}) is expressed by the weighted sum of the expected return of each asset as
\begin{equation}
{\rm E}[R_t]=\sum_{m=1}^Mw_m{\rm E}[r_t^{(m)}],
\label{eq:ERTN}
\end{equation}
where ${\rm E}[{\cdot}]$ denotes the expectation for a random variable.  The covariance matrix of the return of the portfolio is defined by
\begin{equation}
{\Sigma}={\rm E}\left[(\mathbf{r}_t-{\rm E}[\mathbf{r}_t])(\mathbf{r}_t-{\rm E}[\mathbf{r}_t])^{\rm T}\right],
\label{eq:COVMT}
\end{equation}
where components of $\mathbf{r}_t$ are the return of each asset and $({\cdot})^{\rm T}$ denotes the transpose of a vector.  With the use of the covariance matrix in Eq.~(\ref{eq:COVMT}), the variance of the portfolio is obtained as
\begin{equation}
{\sigma}^2=\mathbf{w}^{\rm T}{\Sigma}\mathbf{w}
\label{eq:CONPRT}
\end{equation}
with $\mathbf{w}$ being a weight coefficient vector.  The MV optimized portfolio with expected return ${\mu}$ is realized as the solution of the minimization for ${\sigma}^2$ in Eq.~(\ref{eq:CONPRT}) subject to $\mathbf{w}^{\rm T}\mathbf{r}_t={\mu}$.  Also, constraints for the weight coefficients can be added to the objective function as a Lagrangian form with multipliers. 
\subsection{Risk parity portfolio}
It has been pointed out that the asset classes of the MV portfolio are not fully allocated.  To disperse the risk contributions of portfolios, risk parity (RP) portfolio constructions have been proposed.  Based on the idea of the RP portfolio construction, a measure of risk contribution was introduced.  \par
The risk contribution of the $m$-th asset is derived from the variance of the RP portfolio as follows:
\begin{eqnarray}
{\sigma}_{m}
&=&w_m\frac{{\partial}{\sigma}}{{\partial}w_m} \\
&=&\frac{({\Sigma}\mathbf{w})_m}{\sqrt{\mathbf{w}^{\rm T}{\Sigma}\mathbf{w}}},
\label{eq:RISKCNT}
\end{eqnarray}
where $({\Sigma}\mathbf{w})_m$ denotes the $m$-th component of ${\Sigma}\mathbf{w}$.  Equal risk contribution for the RP portfolio requires that all the risk contributions have the same value, whereby the weight coefficients of the portfolio are determined by optimization as follows:
\begin{equation}
\argmin_{\mathbf{w}}\sum_{m=1}^M\left[w_m-\frac{{\sigma}}{({\Sigma}\mathbf{w})_mM}\right]
\label{eq:OPTRP}
\end{equation}
This portfolio construction enables one to obtain equally allocated assets.  In addition various subclasses of the RP portfolio construction have been proposed.  For instance, the return weighted RP portfolio construction was developed to improve the performance of the equally risk allocated RP portfolio \cite{Baltas-Jessop-Jones-Zhang}. 
\subsection{Risk diversification}
In general, the origin of the risk of assets seems to be entangled.  Namely, the covariance matrix of the return of portfolios contains non-diagonal components and thus the pair of assets exhibit a linear correlation.  To unravel the entangled risks, the PCA has been incorporated into the portfolio constructions \cite{Meucci}.   \par
The covariance matrix of the return of portfolios can be transformed into a diagonal matrix by an appropriate orthogonal matrix since all of the eigenspaces are mutually independent.  The eigenvalues of the covariance matrix introduce a probability distribution of risk contribution.  Thus the entropy with respect to the probability distribution is defined and is employed as the objective function of the MRD portfolio construction.  The origin of risk is expected to be decomposed on the principal axes of the covariance matrix.
\section{Complex valued risk diversification portfolio construction}
\label{sec:CVRD}
As has been reviewed in the previous section, almost all of the portfolio construction methods utilize the covariance matrix to estimate the risk of the portfolios as the objective functions of the optimization.  The covariance matrix of a random vector contains autocorrelations of pairs of vector components.  Thus one can extract stationary information of the random vector from the corresponding covariance matrix.  However, in general, the price of an asset exhibits non-stationary random fluctuations.  Hence it is necessary to utilize dynamic information of  the fluctuations of the assets to accurately estimate the risk of the portfolios. \par
In order to incorporate the dynamics of the price of the assets into the portfolio constructions, we apply the method of EOFs to the timeseries of the return of the assets.  This portfolio construction method utilizes a complex valued timeseries and the corresponding covariance matrix whereby we name this method a complex valued risk diversification (CVRD) portfolio construction. \par
The Hilbert transform of a timeseries $x(t)$ on $t{\in}[0,{\infty})$ is defined by
\begin{equation}
\mathcal{H}[x(t)]=\frac{1}{\pi}\int_0^{\infty}\frac{x({\tau})}{t-{\tau}}d{\tau},
\label{eq:HT}
\end{equation}
where the improper integral is understood in the sense of principal value\cite{Bracewell}.  In practice, empirical timeseries are recoded at a certain sampling rate ${\Delta}t$, which introduces discrete time $t_n=n{\Delta}t$ with $n$ being integer.  The Hilbert transform for a discrete timeseries is given by
\begin{equation}
\mathcal{H}_D[x_k]=-{\rm i\;}{\rm sgn}\left(k-\frac{N}{2}\right)\sum_{n=0}^{N-1}x_n{\rm e}^{{\rm i}\frac{2{\pi}n}{N}},
\label{eq:DHT}
\end{equation}
where ${\rm sgn}(\cdot)$ is the sign function \cite{Kak}.  Here we apply the Hilbert transform in Eq.~(\ref{eq:DHT}) to the return of the portfolio in Eq.~(\ref{eq:RTNPRT}) and then obtain the analytic signal as
\begin{equation}
z_t=r_t+{\rm i}\mathcal{H}_D[r_t].
\label{eq:DAS}
\end{equation}
As with the PCA for real valued time series, the analytic signal, $z_t\;(0{\leq}t{\leq}T)$, provides a complex valued covariance matrix defined as 
\begin{equation}
C_z=\frac{1}{T+1}\sum_{t=0}^{T}z_tz_t^*
\label{eq:CVCM}
\end{equation}
with $z_t^*$ being the transjugate of $z_t$.  Since $C_z$ in Eq.~(\ref{eq:CVCM}) is a positive definite Hermitian matrix, the corresponding eigenvalues are positive real values including zeros.  An unitary matrix $U$, which consists of eigenvectors of $C_z$, transforms $C_z$ as
\begin{equation}
UC_zU^*={\Lambda},
\label{eq:UFT}
\end{equation}
where ${\Lambda}$ is the orthogonal matrix with respect to the set of eigenvalues $\{{\lambda}_n\}$ of $C_z$, which is arranged in descending order.  The weight coefficient vector $\bf{w}$, at the same time, is transformed into $\tilde{\bf{w}}=U\bf{w}$.  The contribution of the eigenvector is introduced as
\begin{equation}
v_m=\tilde{w}_m^2{\lambda}_m
\end{equation}
with $\tilde{w}_m$ being the $m$-th component of the transformed weight coefficient $\bf{\tilde{w}}$, and the probability distribution for $v_m$ is defined by
\begin{equation}
p_m=\frac{v_m}{\sum_{m=1}^Mv_m}.
\label{eq:PD}
\end{equation}
From the probability distribution in Eq.~(\ref{eq:PD}), the corresponding entropy can be introduced as
\begin{equation}
H=-\sum_{m=1}^Mp_m\log{p_m}.
\label{eq:ENT}
\end{equation}
In general, weight coefficients of portfolios are constrained based on trading strategies.  Thus we construct a Lagrangian function with the aid of the entropy in Eq.~(\ref{eq:ENT}) and constraint functions for weight coefficients as 
\begin{equation}
L=H-\sum_{l=1}^L{\mu}_lg_l(\bf{\tilde{w}}),
\label{eq:LGF}
\end{equation}
where ${\mu}_l$ is a Lagrange multiplier and $g_l(\cdot)$ is a constraint function.  Optimizing $L$ in Eq.~(\ref{eq:LGF}) with respect to $\bf{w}$ gives the weight coefficients of the expected CVRD portfolio.
\section{Result}
\label{sec:RSLT}
In this section, we test the performance of the CVRD portfolio construction by comparing it with the RP and MRD portfolio constructions.  As a test dataset, we selected bonds, commodities, indexes and swaps during May 2000 to April 2017.  The descriptive statistics of the dataset are shown in table~\ref{tabl:statistics}.  \par
\begin{figure}
\begin{center}
\includegraphics[scale=0.9]{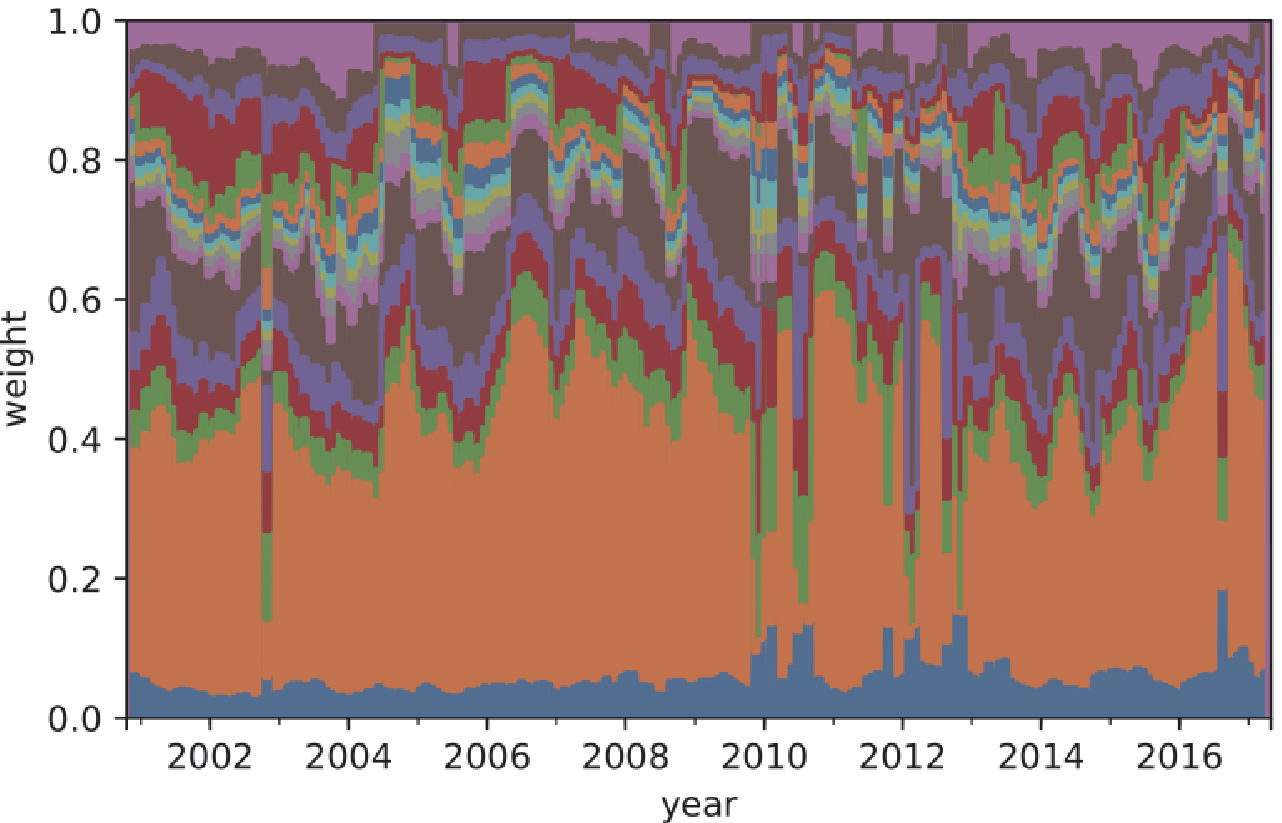}
\caption{\label{fig:WCRP}The allocation of the assets in the RP portfolio construction.}
\end{center}
\end{figure}
\begin{figure}
\begin{center}
\includegraphics[scale=0.9]{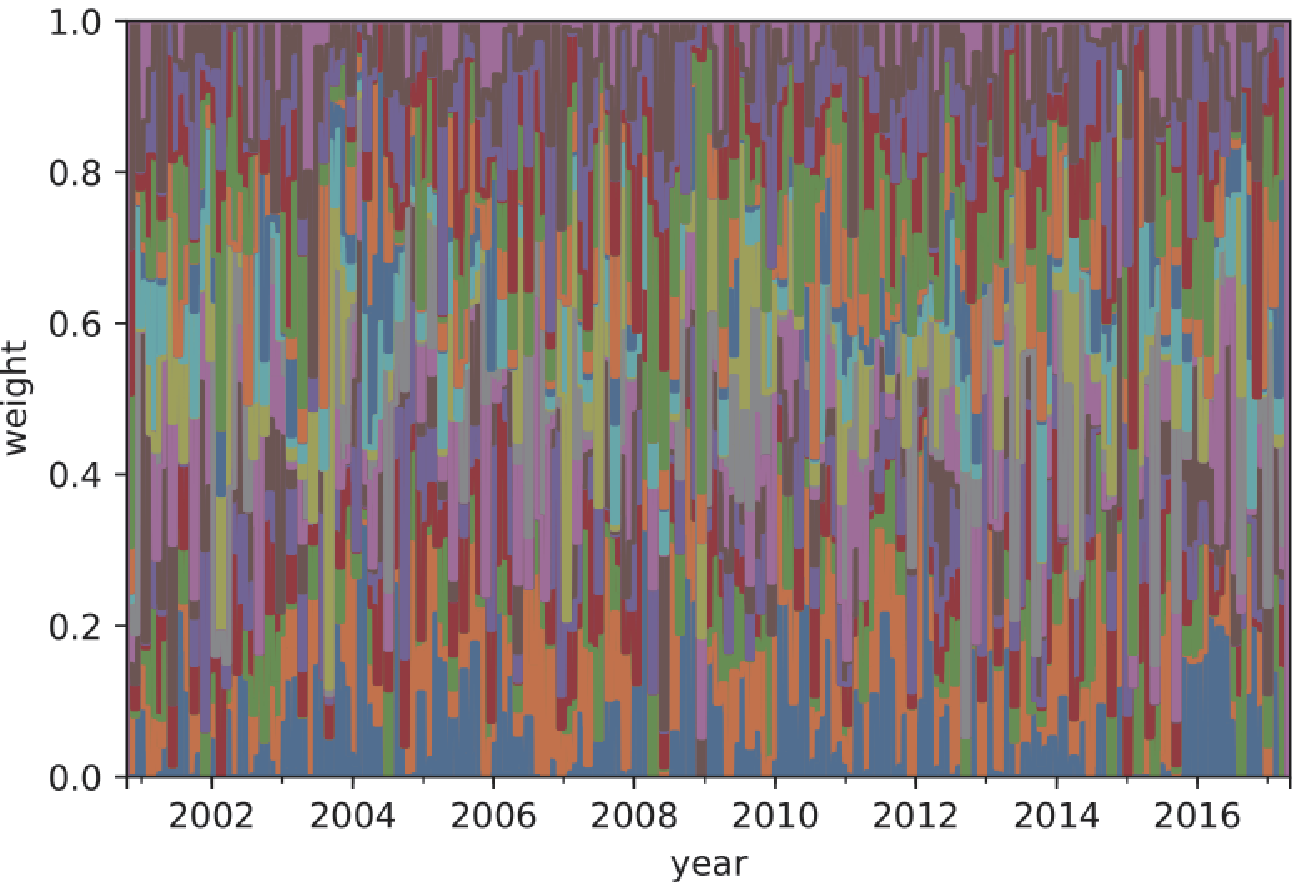}
\caption{\label{fig:WCCVRD}The allocation of the assets in the CVRD portfolio construction.}
\end{center}
\end{figure}

We use the annual return, risk and Sharpe ratio as measures of the performance of the portfolio constructions.  Each portfolio is rebalanced every month by previous data from a year without transaction costs.  Table~\ref{tabl:performance} shows the annual return, risk and Sharpe ratio of the RP, MRD, and CVRD portfolio constructions.  The CVRD portfolio construction outperforms the RD portfolio construction with respect to all of the measures.  The risk of the RP portfolio construction is the lowest because the RP portfolio mainly consists of bonds, which means that the risk contributions of the RP portfolio are strongly biased toward the bonds and thus is not fully dispersed, as is seen in fig.~\ref{fig:WCRP}.  On the other hand, the risk of the assets in the CVRD portfolio construction are well allocated as is shown in fig. \ref{fig:WCCVRD}.   \par
\begin{figure}
\begin{center}
\includegraphics[scale=0.9]{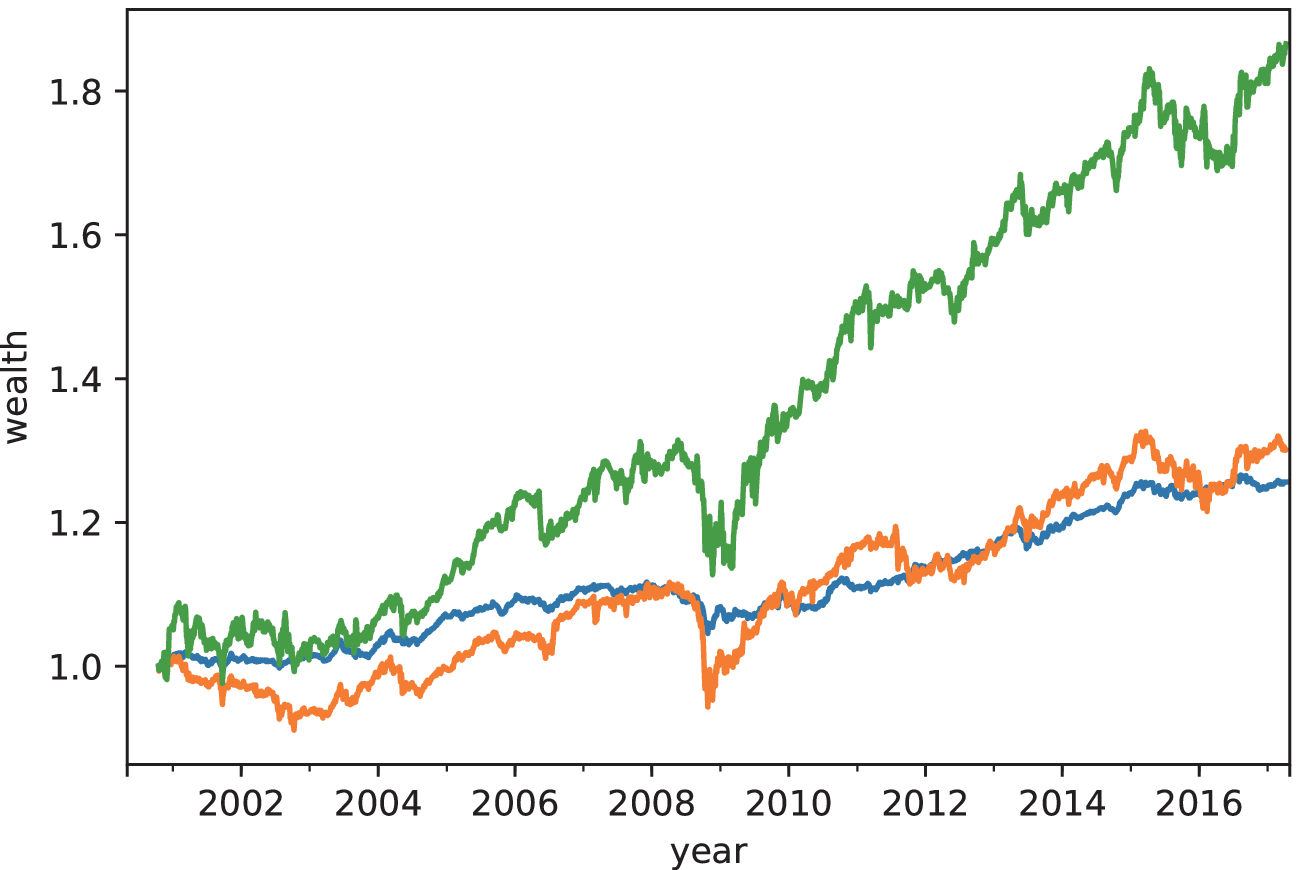}
\caption{\label{fig:Return}The annual returns of the RP (blue), the RD (orange), and the CVRD (green) portfolio constructions.}
\end{center}
\end{figure}

Figure \ref{fig:Return} shows the time sequence of the annual return of the RP, MRD, and CVRD portfolio constructions.  The return of the CVRD portfolio construction is confirmed to outperform that of the RP and MRD portfolio constructions during almost all the periods.  This result seem to be realized by capturing the dynamics of the return of the portfolio with the use of the complex valued PCA.  In other words, the CVRD portfolio construction can appropriately time the rebalancing of the portfolio based on the dynamic properties of the assets.  
%
\begin{table}[htb]
\caption{\label{tabl:statistics} Descriptive statistics of the dataset.}
 \begin{center}
  \begin{tabular}{lcccc} \hline
   & Mean & Std. Dev. & Skewness & Kurtosis \\ \hline
   TY1 Comdty & 0.00661 & 0.465 & -0.284 & 4.56 \\
   XM1 Comdty & 0.000876 & 0.0621 & -0.140 & 1.81 \\
   CN1 Comdty & 0.00935 & 0.444 & -0.222 & 2.87 \\
   RX1 Comdty & 0.01288 & 0.464 & -0.821 & 6.95 \\
   G1 Comdty & 0.00334 & 0.532 & -7.35 & 214 \\
   JB1 Comdty & 0.00414 & 0.283 & -0.571 & 5.78 \\
   SP1 Index & 0.210 & 14.7 & -0.250 & 5.37 \\
   XP1 Index & 0.624 & 46.9 & -0.289 & 5.60 \\
   PT1 Index & 0.0776 & 7.42 & -0.628 & 7.42 \\
   GX1 Index & 1.10 & 90.8 & -0.227 & 4.01 \\
   Z1 Index & 0.166 & 61.3 & -0.223 & 3.61 \\
   NK1 Index & 0.165 & 194 & -0.302 & 5.44 \\
   AUD Curncy & 0.000037 & 0.00641 & -0.361 & 6.11 \\
   CAD Curncy & -0.000026 & 0.00686 & 0.0726 & 2.92 \\
   EUR Curncy & 0.0000400 & 0.00772 & 0.0346 & 1.85 \\
   GBP Curncy & -0.0000600 & 0.00939 & -0.665 & 7.55 \\
   JPY Curncy & 0.000695 & 0.675 & -0.147 & 3.19 \\ \hline
  \end{tabular} 
 \end{center}
\end{table}

\begin{table}[htb]
\caption{\label{tabl:performance} Annual return, risk and, Sharpe ratio of the RP, RD and CVRD portfolio constructions.  Each portfolio is rebalanced every month by previous data for a year.    }
 \begin{center}
  \begin{tabular}{lccc} \hline
     & RP & RD & CVRD \\ \hline 
    Return & 1.340 & 1.621 & 3.816 \\
    Risk & 1.728 & 4.219 & 6.152 \\
    Sharpe Ratio & 0.7756 & 0.384 & 0.620 \\ \hline
  \end{tabular}
 \end{center}
\end{table}
%
%
%
%
%
\section{Conclusion}
\label{sec:CNCL}
Risk diversification for portfolio management is of great interest for both individual and institutional investors.  Indeed, various portfolio construction methods have been developed and employed in both individual and industrial trades.  Nevertheless, almost all of the portfolio constructions fail to account for the dynamic property of the assets.     \par
To utilize the dynamic property of the assets, we introduce the method of EOFs into portfolio constructions.  The Hilbert transform is used to produce the imaginary part of the analytic signal, from which the complex valued covariance matrix obtained.  The PCA for the covariance matrix enables one to estimate the contribution of each principal axis and to obtain the entropy of the risk contribution distribution.  Appropriately constrained optimization methods with respect to the entropy yields the CVRD portfolio construction. \par
The performance of the CVRD portfolio construction was compared with that of the RP and MRD portfolio constructions.  It was confirmed that the annual return of the CVRD portfolio construction outperformed that of the others.  In addition, the risk of the assets in the CVRD portfolio construction was well allocated.  This result verified that the CVRD portfolio construction succeeded in diversifying the risk of the portfolio.  \par
In practice, the time window of estimating the covariance matrix varies on the investors' policy.  Also, the accessible test period depends on the resources of the institution where investor belongs.  Hence the performance of the CVRD portfolio construction seems to vary depending on the size of the time windows and test periods.  Comprehensive investigation for the effect of the time window and the test period is our future work.
%





\end{document}